\documentclass[aps,prd,twocolumn,a4paper]{revtex4}
\usepackage{ulem}
\usepackage{color}
\usepackage{graphicx}
\usepackage{epsfig}
\newcommand{\be}{\begin{equation}}
\newcommand{\ee}{\end{equation}}

\newcommand{\bea}{\begin{eqnarray}}
\newcommand{\eea}{\end{eqnarray}}
\begin{document}
\title{Nonconformal holographic model for D-meson suppression at
energies available at the CERN Large Hadron Collider}
\author{Santosh K. Das$^{1,2,3}$ and Ali Davody$^{4}$}

\affiliation{$^1$ Department of Physics,Yonsei University, Seoul, Korea}
\affiliation{$^2$ Department of Physics and Astronomy, University of Catania,
Via S. Sofia 64, 1-95125 Catania, Italy}
\affiliation{$^3$ Laboratori Nazionali del Sud, INFN-LNS, Via S. Sofia 62, I-95123 Catania, Italy}
\affiliation{$^4$ School of Particles and Accelerators, Institute for research in Fundamental Science
P. O. Box 11365-9161, Tehran, Iran}

\begin{abstract}
The drag force of charm quarks propagating through a thermalized system of
Quark Gluon Plasma (QGP) has been considered  within the framework of both
conformal/non-conformal
Anti de Sitter (AdS)  correspondence. Newly derived Einstein Fluctuation-Dissipation relation has been used to calculate
the heavy flavor diffusion coefficients. Using the drag and diffusion
coefficients as inputs Langevin equation has been solved to study the heavy flavor suppression
factor. It has been shown that within conformal AdS correspondence
the D-meson suppression at LHC energy can be reproduced where as the non-conformal
AdS correspondence fail to reproduce the experimental results. It suggests
collisional loss alone within non-conformal AdS correspondence  can not reproduce the experimental results
and inclusion of radiative loss becomes important.

\vspace{2mm}
\noindent {\bf PACS}: 25.75.-q; 24.85.+p; 05.20.Dd; 12.38.Mh

\end{abstract}
\maketitle

\section{Introduction}
The nuclear collisions at Relativistic Heavy Ion Collider (RHIC)
and the Large Hadron Collider(LHC) energies
are aimed at creating a new state of matter where the bulk properties of the matter
are governed by the light quarks and gluons. Such a
state of matter is called Quark Gluon Plasma (QGP)~\cite{Shuryak:2004cy}.
The study of QGP is a field of great contemporary
interest and the heavy flavors, mainly, charm and bottom quarks, play a
vital role in such studies. This is because heavy quark do not constitute
the bulk part of the system and their thermalization time scale is larger than the light
quarks and gluons and hence heavy quarks can retain the interaction history very effectively.
Therefore, the propagation of heavy quarks through QGP can be treated as the
non-equilibrium heavy quark executing Brownian motion ~\cite{hfr,ahep} in the thermal bath of QGP and
the Langevin equation can be used to study such a system.

In the recent past several attempt has been made to study both
heavy flavor suppression~\cite{phenixe,stare} and elliptic flow~\cite{phenixelat} within the framework of
perturbative QCD~\cite{moore,ko,adil,urs,gossiaux,das,skdas,san,gre,alberico,MBAD,yo,DKS,jeon,bass,hees,dca}.
However it is pointed out that the perturbative
expansion of the charm-quark diffusion coefficient is not well convergent ~\cite{huot}
at the temperature range attainable at RHIC and LHC collisions. Hence, non-perturbative~\cite{rappv2,hvh2,he}
contributions are important to improve heavy quark diffusion.
One possible alternative way to estimate the drag force is the
gauge/string duality~\cite{Maldacena:1997re}, namely the conjectured equivalence between
conformal N=4 SYM gauge theory and gravitational theory in Anti de Sitter
space-time i.e. AdS/CFT. Some attempts have been made in this direction to study heavy
flavor suppression. Within this AdS/CFT model RHIC results has been
reproduced well~\cite{hirano}  whereas
a few other attempts~\cite{gyu,hor} suggest that AdS/CFT under predicts
recent ALICE results~\cite{alice}.
The non-zero value of bulk viscosity obtained form  lattice QCD calculations~\cite{mey} indicate
that at the temperature range relevant of
RHIC and LHC collisions the fluid behavior is non-conformal.
Therefore,  it would be interesting to construct a gravitational dual which
captures some of the properties of QCD.
This can be done by breaking the conformal symmetry in the AdS space
and construct AdS/QCD models ~\cite{adsqcd,adsqcd1}. In this paper we have made an
attempt to test these  AdS/QCD models  by studying D-meson suppression at LHC collision
energies.

\section{Langevin equation and holography}
Consider a heavy quark of mass $M$ and energy $E$ passing through QGP at a temperature, $T (<< M)$.
The heavy quark suffers random kicks leading to momentum transfer  $q\sim T$ in a single elastic collision
with the thermal bath. Hence, it requires many collisions to change the heavy quark momentum significantly.
The dynamics of heavy quarks propagating through the QGP can thus be approximated as a succession of
uncorrelated momentum kicks which leads to a Fokker-Planck equation that can be realized  from the
Langevin equation~\cite{BS,das,san}.
\be
{d p_i \over  dt}=-\gamma_D p_i +\xi_i,\;\;\;\;\;\;   <\xi_i(t)\xi_j(t')>=D_{ij}\delta(t-t')
\ee
where $\gamma_D$ is the drag coefficient, $\xi$ is the random force and $D$ is the diffusion coefficient.

\subsection{Drag and diffusion coefficients in conformal holography}
AdS/CFT in its original form, relates $N=4$ SYM gauge theory on four-dimensional space time to the
IIB string theory on $AdS_5\times S^5$ background, where the conformal
symmetry of SYM gauge theory is realized in the conformal isometry of dual
metric \cite{Maldacena:1997re}. This correspondence can also be generalized to
finite temperature, where the
 space-time dual  to  $N=4$ SYM plasma with temperature $T$ is a black-hole AdS with Hawking
temperature $T$   \cite{Witten:1998qj}. The  metric of AdS black hole is
\be\label{adsmetric}
ds^2={1\over r^2}({dr^2\over  f(r)}-f(r)dr^2+d\overrightarrow{x}^2)\;\;\;\;\;\;\;\;\;f(r)=1-({r_h\over r})^4
\ee
where $r_h=1/(\pi T)$ is the horizon of the black hole. According to  the standard AdS/CFT prescription, the energy-momentum tensor
of dual theory is encoded in  behavior of metric near the boundary. Using this dictionary we find that metric (\ref{adsmetric})  is dual
to a  plasma with conformal equation of state, $e=3p$.

By studying the dynamics of trailing string
in this background \cite{Herzog:2006gh,Gubser:2006bz}, it has been shown that the drag force exerted
on a moving quark in a static N=4 SYM plasma is given by
\be\label{dragconf}
F_{conf}={dp\over dt}= -{\pi\sqrt\lambda\over2}T_{SYM}^2{p\over M} \equiv -\Gamma_{conformal} \;\;p
\ee

Also by investigating the fluctuations around the classical string configuration \cite{Gubser:2006nz,CasalderreySolana:2007qw},
one finds the transverse  diffusion coefficient as follows

\bea\label{confdiff}
&D&=\pi\sqrt\lambda\; \gamma^{1\over 2}\;T_{SYM}^3     
\eea
where $\gamma$ is the Lorentz factor,$\gamma=1/\sqrt{1-v^2}$.

Using (\ref{dragconf}) and (\ref{confdiff}) we find the following "modified Einstein relation"
between drag and diffusion coefficient \cite{HoyosBadajoz:2009pv,Gursoy:2010aa}

\bea
&D&=2 M \, T_{SYM} \sqrt\gamma\; \Gamma_{conformal}
\eea
In terms of world-sheet temperature, $T_s=T/\sqrt \gamma$, the above equation takes the following form
\bea\label{mode}
&D&=2 E \, T_s\; \Gamma_{conf}
\eea
this is the usual Einstein relation for a relativistic particle moving in a thermal bath with temperature $T_s$.
So the  world-sheet temperature, $T_s=T/\sqrt \gamma$,  is the effective temperature for a quark moving in a static plasma  \cite{Gursoy:2010aa}. \\

In order to apply  $N=4$ SYM results to  QCD, we use alternative scheme introduced in\cite{Gubser:2006qh}\footnote{Using this scheme it has been shown \cite{hirano}  that AdS/CFT predictions  lead to reasonable results at RHIC energy }.
According to this proposal, one equates the energy density of QCD and SYM, which leads to $T_{SYM}=T_{QCD}/{3^\frac{1}{4}}$.
Also by comparing string prediction for quark-antiquark potential with lattice gauge theory we find that, $3.5<\lambda<8$ \cite{Gubser:2006qh}.
Therefore in terms of QCD temperature and coupling we have

\bea\label{confdrag}
&\Gamma_{conf}&=\alpha {T_{QCD}^2\over M}\\
&D_{conf}&={2\alpha\over 3^{1\over4}}\gamma^{1\over 2}T_{QCD}^3    
\eea

where $\alpha={\pi\sqrt\lambda\over 2\sqrt3}=2.1\pm0.5$.

\subsection{Drag and diffusion coefficients in non-conformal holography}
In the previous section we have described different ways of evaluating the drag and diffusion
coefficients of SYM plasma.
However SYM and QCD have different properties: equation of state, phase transition, symmetries, etc.
In particular  SYM plasma is a conformal fluid with vanishing bulk viscosity. On the other hand, QGP looks like a
conformal fluid at very high enough temperature, $T>>T_c$.
So it would be interesting to construct a gravitational dual which
captures some of the properties  of QCD \cite{Gursoy:2007cb,Gursoy:2007er,Gubser:2008ny}.
To do so, we have to break the conformal symmetry of AdS space. In   \cite{Gursoy:2007cb,Gursoy:2007er}
a  5-dimensional non-conformal gravitational model dual to QCD  is proposed,
where the non-trivial profile of dilation breaks down the conformal symmetry. The action of 5-dimensional Einstein-dilation  is given by
\be
S=-{1\over 16 \pi G_5}\int d^5x \sqrt{-g}\,(R-{4\over 3}(\partial\phi)^2+V(\phi))
\ee
where $G_5$ is the five-dimensional Newton  constant. By choosing a suitable  scalar potential one can mimics
 the QCD equation of state and other thermal properties.
We choose the suggested potential in \cite{Gursoy:2009kk}:

\be
V(\lambda)=\frac{12}{l^2}\{ 1+ V_0 \lambda + V_1 \lambda^{4\over 3}[\ln(1+V_2 \lambda^{4\over3}+V_3 \lambda^2)]^{1\over 2}\}
\ee

with

\bea
V_0=\frac{8}{9}\beta_0,\;\;\;V_2=\beta_0^4 \left(\frac{23+36{\beta_1 \over \beta_0^2}}{81 V_1}\right)^2,\;\; \nonumber\\
\beta_0=\frac{22}{3(4\pi)^2},\;\;\beta_1=\frac{51}{121}\beta_0^2.
\eea

\begin{figure}[ht]
\begin{center}
\includegraphics[scale=0.35, clip=true]{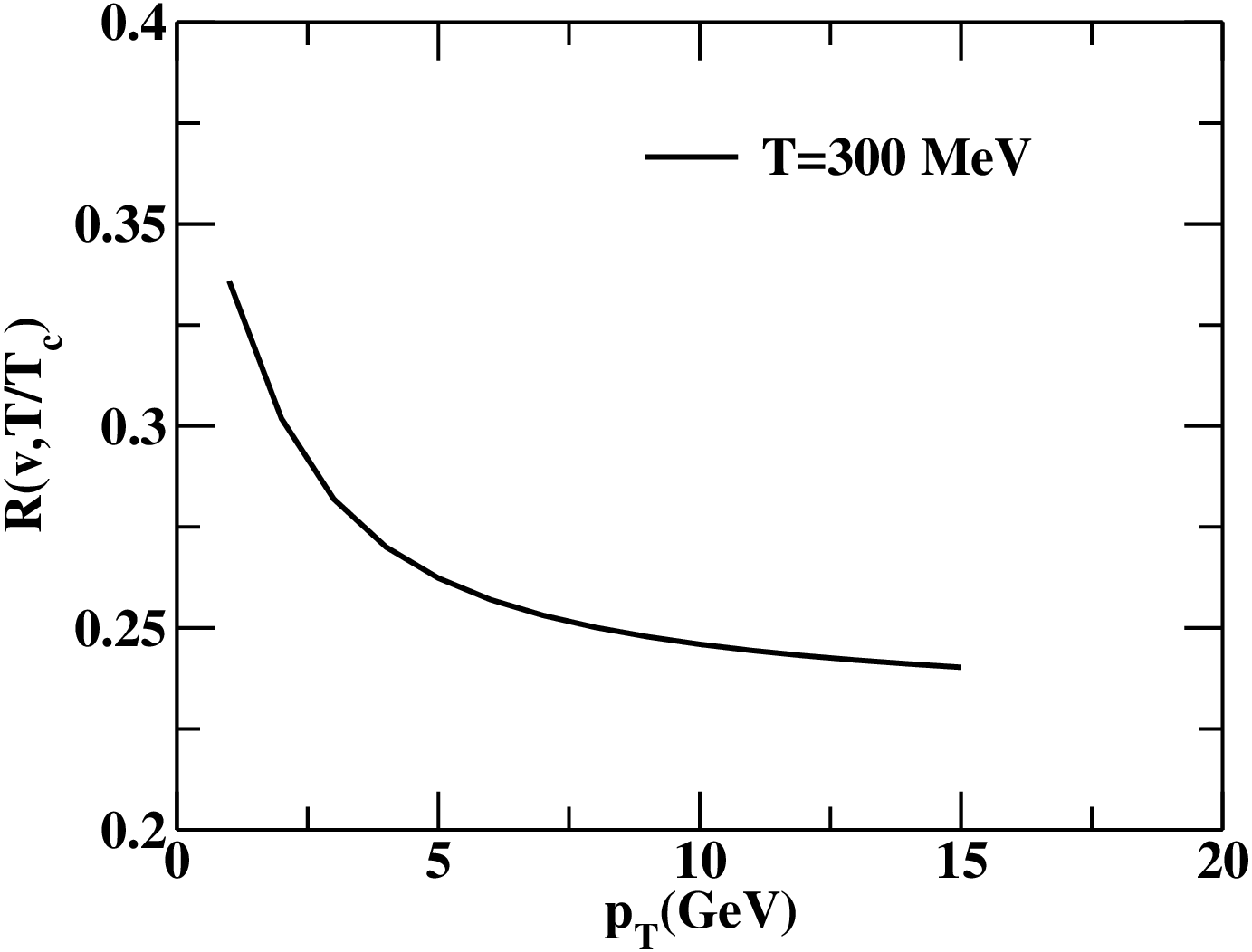}
\caption{Variation of $R(v,T/T_c)$ as a
function of momentum  $p_T$.  }
\label{fig1}
\end{center}
\end{figure}

\begin{figure}[ht]
\begin{center}
\includegraphics[scale=0.35, clip=true]{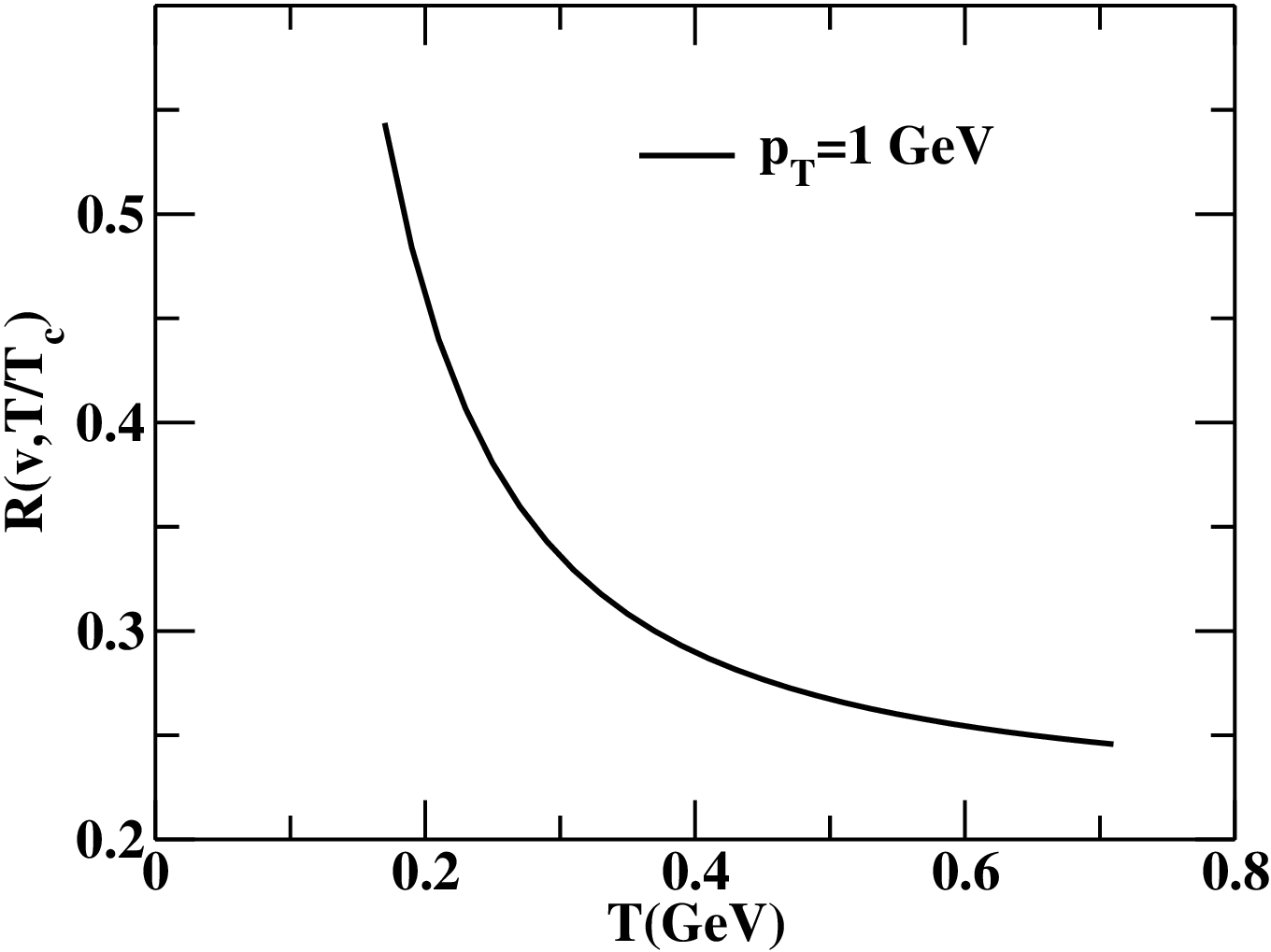}
\caption{Variation of $R(v,T/T_c)$ as a
function of temperature $T$.  }
\label{fig2}
\end{center}
\end{figure}

It has been shown \cite{Gursoy:2009kk} that this potential reproduces the lattice EOS and velocity of sound.  Drag force in this model  is calculated in  \cite{Gursoy:2009kk}:

\be
F_{non-conf}={- v\; e^{2A_s(r_s)}\over 2 \pi l_s^2}\equiv \Gamma_{non-conf} \, p
\ee
where $v$ is the speed of quark, $r_s$ is the world-sheet horizon and $A_s(r_s(v))$ is conformal factor of metric in string frame evaluated at the world-sheet horizon. Here $l_s$ is a \emph{free parameter} which can be fixed by matching the string tension to the
 string tension derived from the lattice QCD calculations and is given by \cite{Gursoy:2009kk}
\be
l_s\simeq \;0.15 \, l
\ee
Note that unlike conformal case, non-conformal  drag coefficient is  velocity dependent through $r_s$.
It is useful to study the ration  of drag force in non-conformal holography to  the conformal case:
\be
{F_{non-conf}\over F_{conf}}={\Gamma_{non-conf}\over \Gamma_{conf}}={2.1\over\alpha}\;R\,(v,{T\over T_c})
\ee
where $R$ is a function of temperature and velocity of quark. We reproduced this function in  Fig.\ref{fig1} and
Fig.\ref{fig2} for completeness  \cite{Gursoy:2009kk} \footnote{Note that $T\over T_c$ is scheme independent}\footnote{we have checked that our calculations correctly reproduce the results of \cite{Gursoy:2009kk}} .
If we take $l_s$ as a free parameter, then this relation takes the following form :


\bea\label{nonconfdrag}
{\Gamma_{non-conf}}=({0.15\; l\over l_s})^2\;{2.1\over\alpha}\;R\,(v,{T\over T_c})\;{\Gamma_{conf}} \nonumber\\
=({0.15 \; l \over l_s})^2 \;\;\;2.1\;R\,(v,{T\over T_c}) {T_{QCD}^2\over M}
\eea

The modified Einstein relation (\ref{mode}) for non-conformal becomes

\bea\label{moden}
&D_{non-conf}&=2 E \, T_{s,non-conf}\; \Gamma_{non-conf}
\eea

where $ T_{s,non-conf}$ is the world-sheet temperature in non-conformal case. In terms of the ratio of world-sheet temperature in non-conformal to conformal case, $G(v,{T\over T_c})={ T_{s,non-conf}\over  T_{s,conf}}$, the above equation takes the following form
\cite{Gursoy:2010aa}

\bea
D_{non-conf}=2 E \; T_{s,conf} \;G(v,{T\over T_c}) \;\Gamma_{non-conf} \nonumber \\
={2 M \over 3^{\frac{1}{4}}} \sqrt\gamma\; T_{QCD} \;G(v,{T\over T_c}) \;\Gamma_{non-conf}
\eea


There are several limits on AdS/CFT results discussed here. In \cite{Abbasi:2012qz}
 the effects of hydrodynamic expansion of QGP on drag force exerted on a moving quark have been studied.
It was shown that there is an upper bound for velocity of quark ($v_{bound} \approx 0.98$) such that below this bound,
drag force acting on the quark is just the localized version of static plasma (replacing the temperature in
drag formula of static plasma by instantaneous temperature of QGP). On the other hand, for a fast quark with a
velocity bigger than the above bound (for charm quark this bound in velocity corresponds to around 10GeV in energy)
drag force is not a local function of the medium variables. Thus, for $p_T\gg 10GeV $ the local approximation
is not valid  due to hydrodynamic expansion.

Also it has been shown in \cite{Gursoy:2010aa} that for $p_T> 10GeV $ the white noise approximation in
Langevin equation breaks down. Above
this bound, one needs the full frequency-dependent correlators to study the diffusion process \cite{Kiritsis:2011bw}.

\section{Initial condition and space time evolution}
After obtaining the drag and diffusion coefficients from the conformal and
non-conformal holography, we need the initial heavy quark momentum distributions
for solving the Langevin equation. In the present work, the $p_T$ distribution of
charm quarks in pp collisions have been generated using the POWHEG~\cite{pow} code, implementing
pQCD at NLO. It should me mentioned here that the $p_T$ distribution of
charm quarks in pp collisions generated using POWHEG can reproduce the experimental results~\cite{alberico,expp}.
With this initial heavy quarks momentum distribution, the Langevin equation
has been solved . We convolute the solution with the fragmentation functions of the
heavy quarks to obtain the $p_T$ distribution of D mesons. For heavy quark fragmentation,
we use the Peterson function~\cite{setiii}. Experimental data (pp collisions) on the electron spectra originated from the
decays of the heavy mesons can be described if Peterson fragmentation is applied to the POWHEG output.
This has been studied in Ref.~\cite{alberico}.

The experimental interest is the nuclear suppression factor $(R_{AA})$, defined as
\be
R_{AA}(p_T)=\frac{\frac{dN}{d^2p_Tdy}^{\mathrm Au+Au}}
{N_{\mathrm coll}\times\frac{dN}{d^2p_Tdy}^{\mathrm p+p}}
\label{raa10}
\ee
a ratio that summarizes the deviation
from what would be obtained if the nucleus-nucleus collision is an
incoherent superposition of nucleon-nucleon collisions.
In Eq.~\ref{raa10}
$N_{\mathrm coll}$ stands for the number of nucleon-nucleon
interactions in a nucleus-nucleus collision.
In the present scenario the variation of temperature with time
is governed by the the equation of state (EOS) or velocity of sound of the
thermalized system undergoing hydrodynamic expansion.
Hence, $(R_{AA})$ is sensitive to velocity of sound.

The system formed in nuclear collisions at relativistic energies evolves dynamically from
the initial QGP state at temperature, $T_i$ to the quark-hadron transition temperature, $T_c$.
Boost invariance Bjorken~\cite{bjorken} model
has been used for the space time description of the QGP.
It is expected that
the system formed in nuclear collisions at RHIC and LHC energy in the central rapidity
region is almost free from net baryon
density. Therefore, the equation governing the conservation of net baryon number need not be
considered in the present case.

The total amount of energy dissipated in the system by the charm quarks
depends on the number of interaction it undergoes {\it i.e.} on the
path length ($L$) it traverses within the medium. The value of
$L$ in turn depends on the spatial co-ordinates, $(r,\phi)$ of
the point of creation of the charm quark. The probability, $P(r,\phi)$
that a charm quark is created at $(r,\phi)$ depends on the number of binary
collisions at that point. $P(r,\phi)$ is given by:
\be
P(r,\phi)=\frac{2}{\pi R^2}(1-\frac{r^2}{R^2})\theta(R-r)
\label{prphi}
\ee
where $R$ is the nuclear radius. It should be mentioned here
that the expression in Eq.~(\ref{prphi}) is an approximation for the
collisions with zero impact parameter. In obtaining the above expression for $P(r,\phi)$
spherical geometry has been assumed, therefore, it is better applicable for central
collisions.  The charm quark created at $(r,\phi)$ in the transverse plane of the
medium will propagate a path length, $L$ given by
$L=\sqrt{R^2-r^2sin^2\phi}-rcos\phi$.
The geometric averaging has been performed for the
the drag and diffusion coefficients along the path length.
The initial temperature ($T_i$) and thermalization time ($\tau_i$)
of the background QGP are constrained by the
following equation:
\begin{equation}
T_i^{3}\tau_i \approx \frac{2\pi^4}{45\zeta(3)}\frac{1}{4a_{eff}}\frac{1}
{\pi R^2}\frac{dN}{dy}.
\label{eq4}
\end{equation}
where ($dN/dy$) is the measured all hadronic multiplicity,
$\zeta(3)$  is the Riemann zeta function and  $a_{eff}=\pi^2g_{eff}/90$ where
$g_{eff}$ ($=2\times 8+ 7\times 2\times 2\times 3\times N_F/8$) is the
degeneracy of quarks and gluons in QGP, $N_F$=number of flavors.
We use the measured total hadronic multiplicity at
central rapidity, $dN/dy \approx 1100$ for RHIC and $dN/dy \approx 2400$ for LHC energies.
Eq.~\ref{eq4} works in absence of viscous loss where the time
reversal symmetry of the system is valid.
Initial conditions for the LHC and RHIC energies has been taken from Ref~\cite{vic}
and Ref~\cite{alberico} respectively.

\section{Results}

\begin{figure}[ht]
\begin{center}
\includegraphics[scale=0.35, clip=true]{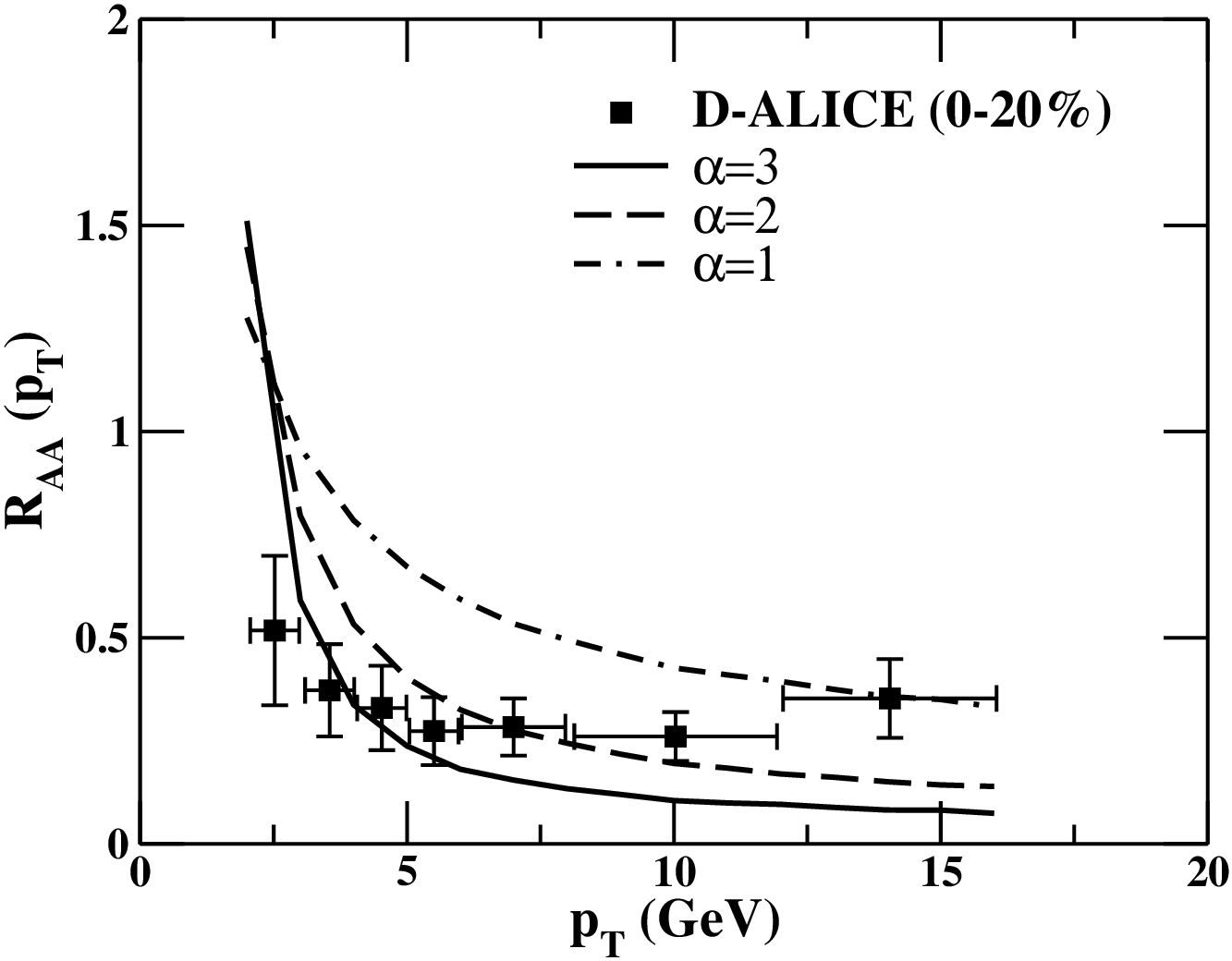}
\caption{Variation of $R_{AA}$ as a
function of momentum  $p_T$ for D meson at ALICE within conformal model.
Experimental data taken from ~\cite{alice}. Although we have present
the results here upto $p_T\sim 15$ GeV, it should be mentioned here that
the white noise approximation in the Langevin equation is not
valid beyond $p_T=10$ GeV. In this backdrop the theoretical results should be taken.}
\label{fig3}
\end{center}
\end{figure}

\begin{figure}[ht]
\begin{center}
\includegraphics[scale=0.35,clip=true ]{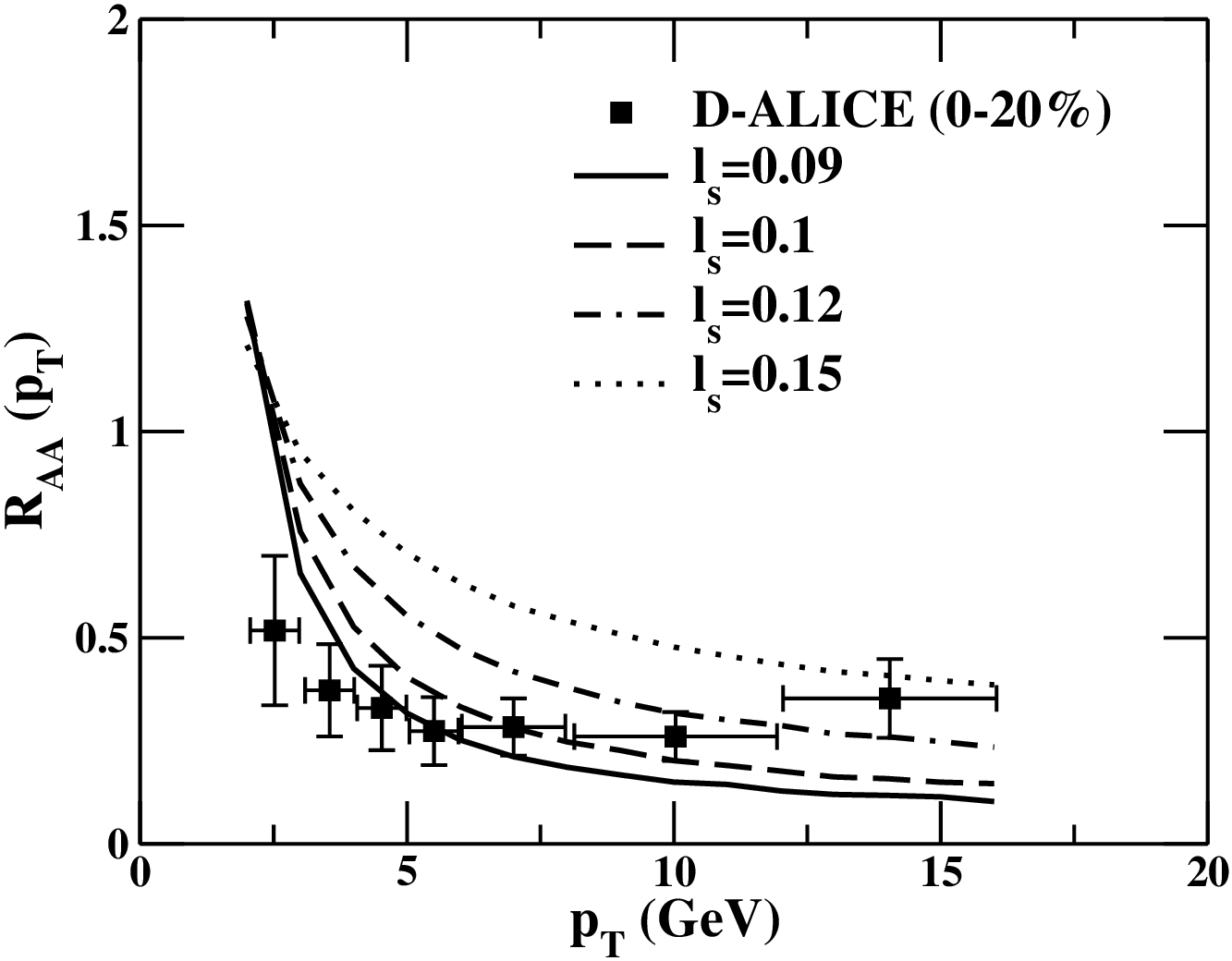}
\caption{Variation of $R_{AA}$ as a
function of momentum  $p_T$ for D meson at ALICE within non-conformal model.
Experimental data taken from ~\cite{alice}. }
\label{fig4}
\end{center}
\end{figure}

The ratio of interaction obtained from non-conformal to conformal
is displayed in Fig~\ref{fig1} with respect to $p_T$. It is observed that
the non-conformal drag force reduced by a factor 3-4 time than the conformal case.
The momentum dependent is also weak as shown in Fig~\ref{fig1}. Similarly in Fig~\ref{fig2} the variation of
the ratio from non-conformal to conformal is plotted with respect to $T$ for fixed $p_T$. It is found
that non-conformal drag force reduced by a factor 2-4 time than the conformal. We use $T_c$=170 MeV.

With the formalism discussed above the result for $(R_{AA})$ are shown
in Fig~\ref{fig3} for the conformal holography. It is found that the
ALICE data can be explained reasonably well for $\alpha=2$. For $\alpha=3$ we under predicts
the experimental data. Here it may be mentioned that
within the conformal holographic model the RHIC results was
explained reasonable well for $\alpha=2-3$~\cite{hirano} in Langevin dynamics.
However the conformal holography model based on HQ energy loss under predict the recent ALICE data~\cite{alice}
presented in ref~\cite{gyu,hor}. As the conformal
results are aways from reality, in a very first attempt we are implemented
the con-conformal results with the Langvine equation to study the D-meson
suppression at LHC energy.

\begin{figure}[ht]
\begin{center}
\includegraphics[scale=0.35, clip=true]{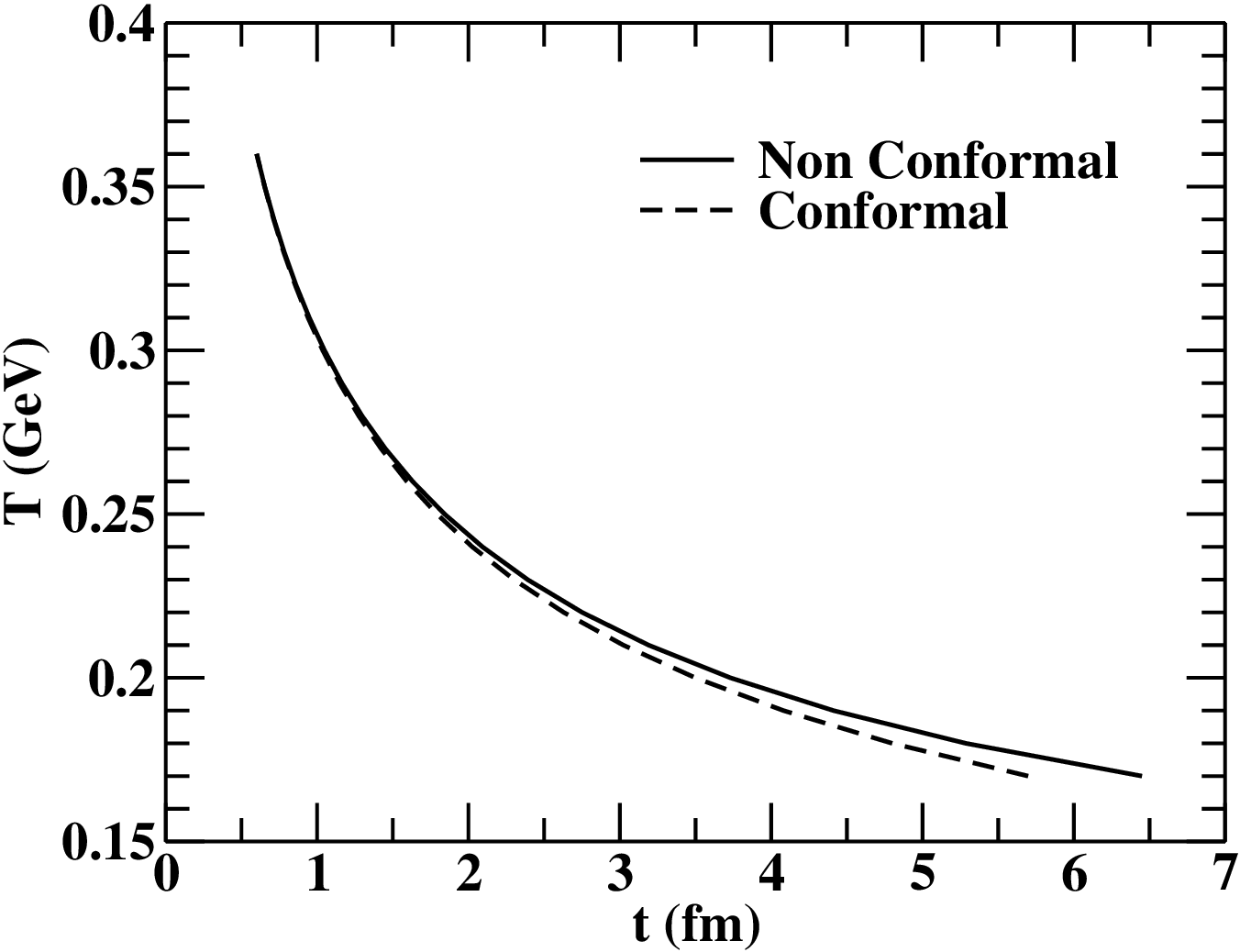}
\caption{Time evolution of the temperature
for both the conformal and non conformal scenarios}
\label{fig5}
\end{center}
\end{figure}

In Fig~\ref{fig4}
the variation of $(R_{AA})$ has been shown as a function of $p_T$ for various value
of $l_s$ within the non-conformal holography . It is found that non-conformal drag force
over predicts the data for a realistic
value of $l_s=0.15$. It is quite expected as the non-conformal drag forces suppressed by a
factor 2-4 as compare to conformal case (in Fig~\ref{fig1} and Fig~\ref{fig2}) depending on
temperature and momentum. Apart from the drag force, the conformal and non-conformal AdS follow
different Einstein relation to have the diffusion coefficients as well as they have different
EOS~\cite{eoss} which indeed affect the RAA~\cite{das,jalam}.
Considering only the collisional
loss within the non-conformal holographic fail to reproduced the experimental data.
The results will improve if the radiative loss from the non-conformal holography will be
taken into account. We may mention, the calculation of radiative energy loss in holography
can be found in Ref.~\cite{rad_con} for conformal case and in Ref.~\cite{rad_non} for non-conformal holography.
In Fig~\ref{fig5} the time evolution of the temperature at the RHIC energy has been shown for both the
conformal and non conformal scenarios. It is found that the time evolution is bit slow
in the non conformal case in comparison with the conformal case and hence the life time
of the QGP is larger for the non conformal case.

\begin{figure}[ht]
\begin{center}
\includegraphics[scale=0.35, clip=true]{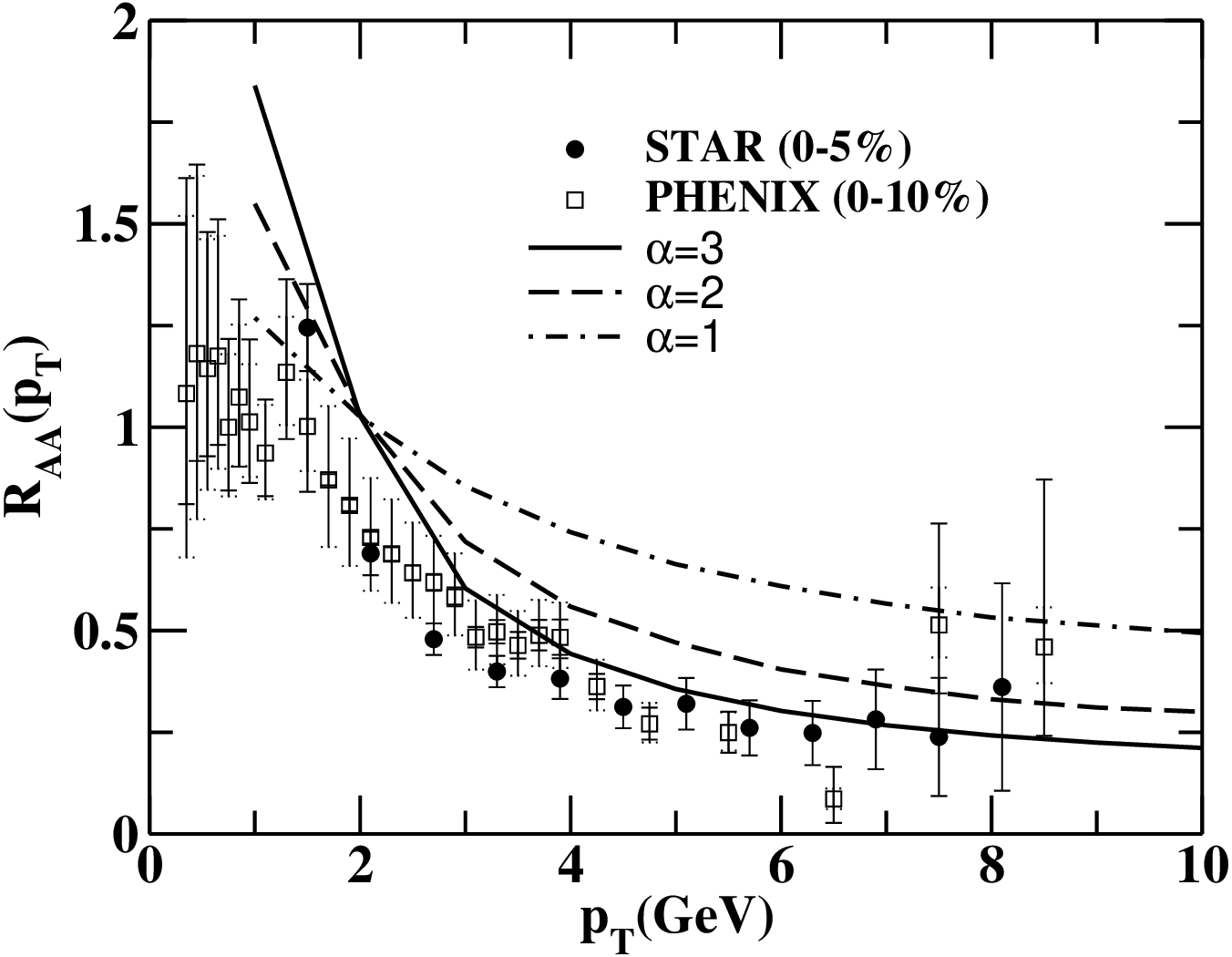}
\caption{ Comparison of $R_{AA}$ obtained within the conformal model  with
the experimental data obtained by STAR and PHENIX collaboration.
Experimental data taken from ~\cite{stare,phenixelat}.}
\label{fig6}
\end{center}
\end{figure}

\begin{figure}[ht]
\begin{center}
\includegraphics[scale=0.35,clip=true]{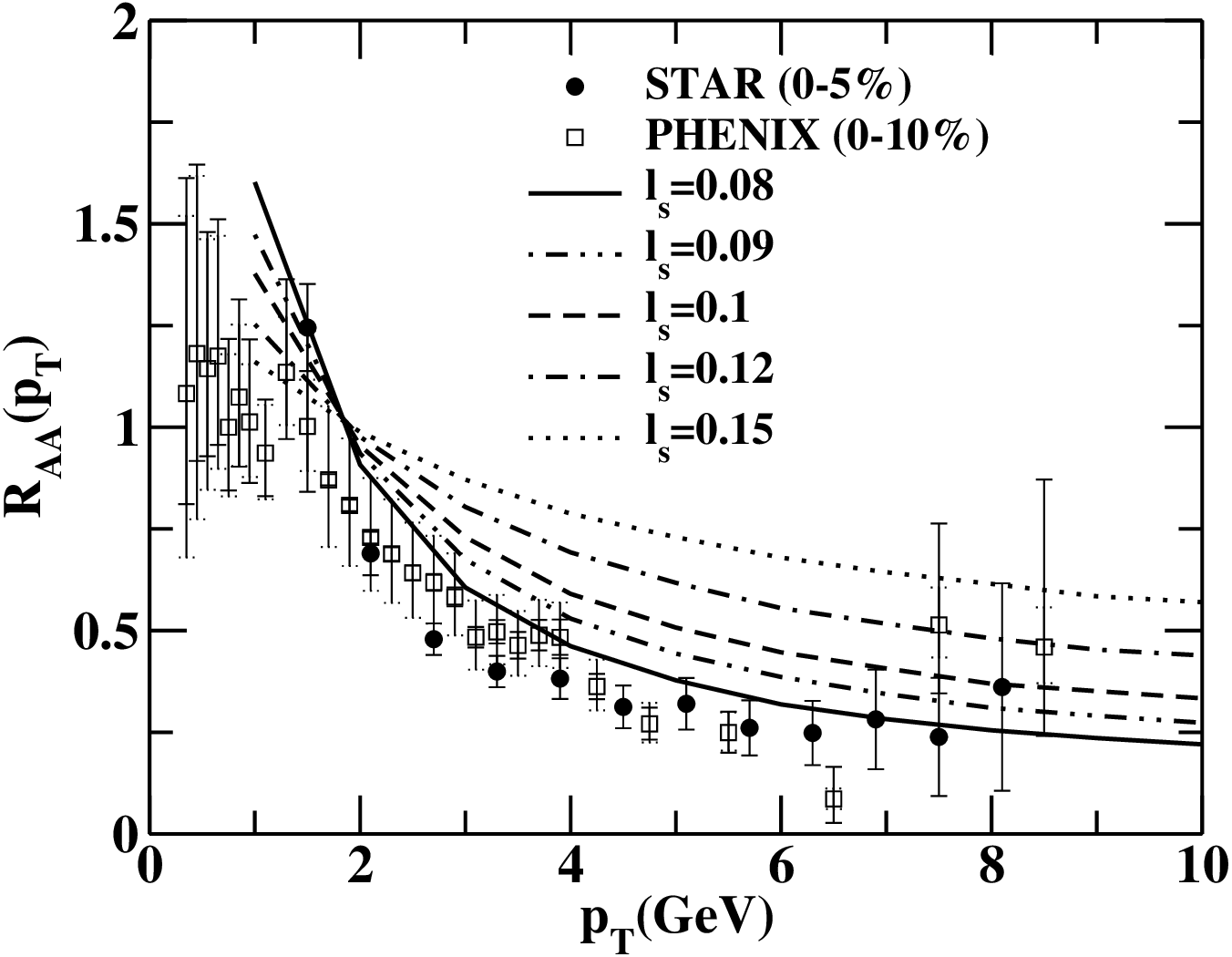}
\caption{Comparison of $R_{AA}$ obtained within the non-conformal model  with
the experimental data obtained by STAR and PHENIX collaboration.
Experimental data taken from ~\cite{stare,phenixelat}.}
\label{fig7}
\end{center}
\end{figure}

The PHENIX and STAR collaborations~\cite{phenixe,stare} have measured the $R_{AA}(p_T)$ of
non-photonic single electrons originating from the decays of mesons containing
both open charm and bottom quarks at RHIC energy. It will be interesting to study the
the RHIC data within the ambit of the present model described above.
The $p_T$ spectra of non-photonic electrons originating from the heavy ion collisions can be obtained as follows
(for details we refer to~\cite{das,san}):
(i) First we obtain the $p_T$ spectra of $D$ and $B$ mesons by  convoluting the solution
of the Langevin equation for the charm and bottom quarks with their respective fragmentation functions as discussed earlier.
(ii) Then we calculate the $p_T$ spectra of the single electrons resulting from the decays of $D$
and $B$ mesons: $D\rightarrow X e \nu$ and $B\rightarrow X e \nu$ respectively.
In the same way, the electron spectrum from the
pp collisions can be obtained from the charm and bottom
quark distributions, which represent the initial conditions
for the solution of the Langevin equation. Theoretical results
obtained within the conformal model contrasted with the experimental
data from RHIC experiments in Fig.~\ref{fig6}. It is found that the
RHIC data can be explained reasonably well within the conformal model for $\alpha=3$.
It can be mentioned that the suppression we are getting for $\alpha=2$ is  less than
the suppression obtained in ~\cite{hirano}. This is may be due to the different initial condition
as well as the uncertainty associated with the addition of electron coming from the decay of D and B mesons~\cite{bass}.
In Fig.~\ref{fig7}, we compare the RHIC data with our results obtained within the non-conformal model.
The results reveal that non-conformal model over predicts the data for the realistic
value of $l_s=0.15$ like the LHC case.

It may be mentioned here that in the present study we are using the Gaussian white noise approximation
to include the collision. According to the recent study ~\cite{Gursoy:2010aa} for the conformal case
white noise approximation will be valid if
\be
T_s>>\eta_D
\ee
where $T_s$ is the world-sheet horizon and $\eta_D$ is the drag force coefficient.
Using $T_s=T/\sqrt(\gamma)$ and the value of the drag coefficient used in the present calculation
leads to the bound on charm quark momentum at $T\sim T_c$ is $p_{max}\sim 10$ GeV and
at $T\sim 2T_c$ is $p_{max}\sim 4.5$ GeV. In this backdrop the theoretical
results should be taken.
The corresponding bound on the bottom quark is about 100 GeV
and 50 GeV at $T\sim T_c$ and
$T\sim 2T_c$ respectively. For non conformal case the momentum bound is much less restrictive
than the conformal case as the non conformal drag coefficient is much smaller than the conformal case.
Moreover the white noise approximation is a good approximation for the non conformal background
than the conformal case.

\section{Summary and conclusions}
In an attempt, we have studied the D-meson suppression at LHC energy within both the conformal and
the non-conformal holographic model. It is observed that the non-conformal holographic model
over predicts the ALICE data where as the data can explained reasonably well for $\alpha=2$
within the conformal holography. This is because the non-conformal drag force
suppressed by a factor of 2-4 compared to the conformal case. The same formalism has
been applied to study the experimental data on nonphotonic
single-electron spectra measured by STAR and PHENIX
collaborations at the highest RHIC energy. The data is well reproduced
within the conformal model for $\alpha=3$ where as the non-conformal holographic model
over predicts the data for the realistic value of $l_s$. We found that, within the conformal
holographic model, RHIC and LHC
data can not be reproduced simultaneously with the same value of $\alpha$.
It is expected that inclusion of the
radiative loss from the non-conformal side will improve the results. Therefore, more systematic studies
are needed from the non-conformal side like inclusion of radiative loss, etc to improve the description
of the experimental results.

\subsection*{Acknowledgements}
S.K.D is thankful to Jane Alam, Andrea Beraudo, Ajay Dash and Marco Ruggieri for very useful discussion.
A.D would like to thank Navid Abbasi,Davood Allahbakhshi,
Elias Kiritsis, Francesco Nitti and Umut Gursoy for useful discussions.
We thank Urs Weidemann for useful discussion during our stay at CERN, theory division, where
this collaboration stated. S. K. D. acknowledges the support by the ERC StG
under the QGPDyn Grant No. 259684.


\begin{thebibliography}{99}
\bibitem{Shuryak:2004cy}
  E.~V.~Shuryak,
  Nucl.\ Phys.\ A {\bf 750} (2005) 64

\bibitem{hfr} R.Rapp  and H van Hees, R. C. Hwa, X. N. Wang (Ed.) Quark Gluon
Plasma 4, 2010, {\it World Scientific}, {\bf 111}    

\bibitem{ahep}S. Majumdar, T. Bhattacharyya and S. K. Das,   Advances in High Energy
Physics, Vol. 2013, Article ID 136587, (2013)

\bibitem{phenixe} S. S. Adler {\it et al.} (PHENIX Collaboration),
Phys. Rev. Lett. {\bf 96}, 032301 (2006).

\bibitem{stare} B. I. Abeleb {\it et al.} (STAR Collaboration), Phys. Rev.
Lett. {\bf 98}, 192301 (2007).

\bibitem{phenixelat} A. Adare {\it et al.} (PHENIX Collaboration),
Phys. Rev. Lett. {\bf 98}, 172301 (2007).

\bibitem{moore} G. D. Moore and D. Teaney, Phys. Rev. C {\bf 71}, 064904 (2005).

\bibitem{ko} C. M. Ko and W. Liu, Nucl. Phys. A {\bf 783}, 23c (2007).

\bibitem{adil} A. Adil and I. Vitev, Phys. Lett. B  {\bf 649}, 139 (2007).

\bibitem{urs} N. Armesto, M. Cacciari, A. Dainese, C.A. Salgado and U.A. Wiedemann, Phys. Lett. B {\bf 637} 362 (2006)

\bibitem{gossiaux}  P. B. Gossiaux and J. Aichelin, Phys. Rev. C
{\bf 78}, 014904 (2008).

\bibitem{das} S. K Das, J. Alam and P. Mohanty,
Phys. Rev. C {\bf 80}, 054916 (2009)

\bibitem{skdas} S. K Das, J. Alam, P. Mohanty and B. Sinha Phys. Rev. C {\bf 81}, 044912
(2010); S. K. Das and J. Alam, PoS {\bf 059}   

\bibitem{san}S. K Das, J. Alam and P. Mohanty,
Phys. Rev. C {\bf 82}, 014908 (2010)

\bibitem{gre} J. Uphoff, O. Fochler, Z. Xu and C. Greiner, Phys. Rev. C {\bf 84} 024908 (2011)

\bibitem{alberico} W. M. Alberico {\it et al.}, Eur.Phys.J.C {\bf 71} 1666 (2011)

\bibitem{MBAD} S. Majumdar, T. Bhattacharyya, J. Alam and S. K. Das, Phys. Rev. C {\bf 84} 044901 (2011)

\bibitem{yo} M. Younus and D. K. Srivastava, J. Phys. G G {\bf 39} 095003 (2012)

\bibitem{DKS} R. Abir, U. Jamil, M. G. Mustafa and D. K. Srivastava,
Phys. Lett. B {\bf 715}, 183 (2012).

\bibitem{jeon} C. Young, B Schenke, S. Jeon and C. Gale, Phys. Rev. C {\bf 86}, 034905 (2012)

\bibitem{bass} S. Cao, G. Qin and S. Bass, J. Phys. G {\bf 40}, 085103 (2013)

\bibitem{hees} T. Lang, H. Hees, J. Steinheimer and M. Bleicher, arXiv:1208.1643 [hep-ph]

\bibitem{dca} S. K. Das, V. Chandra, J. Alam, J. Phys. G {\bf 41} 015102 (2014)

 \bibitem{jalam} S. Majumdar and J. Alam, Phys.Rev. C {\bf 85} (2012) 044918


\bibitem{huot} S. Caron-Huot and G. D. Moore, Phys. Rev. Lett. {\bf 100} 052301 (2008).

\bibitem{rappv2} H. van Hees, V. Greco and R. Rapp, Phys. Rev. C
{\bf 73}, 034913 (2006).

\bibitem{hvh2}  H. van Hees, M. Mannarelli, V. Greco and R. Rapp,
Phys. Rev. Lett. {\bf 100}, 192301 (2008).

\bibitem{he} M. He, R. J. Fries and R. Rapp, Phys. Rev. C {\bf 86}, 014903 (2012)

\bibitem{Gubser:2006bz} S.~S.~Gubser, Phys.\ Rev.\ D {\bf 74}, 126005 (2006)  

\bibitem{Maldacena:1997re}
  J.~M.~Maldacena,
    Adv.\ Theor.\ Math.\ Phys.\  {\bf 2}, 231 (1998)

\bibitem{hirano} Y. Akamatsu, T. Hatsuda and T. Hirano,
Phys. Rev. C {\bf 79}, 054907 (2009) .

\bibitem{gyu} W. A. Horowitz and M. Gyulassy, J.Phys.G {\bf 35} 104152 (2008)

\bibitem{hor} W.A. Horowitz, arXiv: 1108:5876 [hep-ph]

\bibitem{alice} B. Abelev {\it et al.},(ALICE Collaboration), J. High Energy Phys. {\b 09} 112 (2012)

\bibitem{mey} H. B. Meyer, Phys. Rev. Lett. {\bf 100} 162001 (2008)

\bibitem{adsqcd} J. Erlich, E. Katz, D. T. Son and M. A. Stephanov, Phys. Rev. Lett. {\bf 95}, 261602 (2005).

\bibitem{adsqcd1} L. D. Rold and A. Pomarol, Nucl. Phys. B {\bf 721}, 79 (2005).

\bibitem{BS} B. Svetitsky, Phys. Rev. D {\bf 37}, 2484( 1988).

\bibitem{Witten:1998qj}
  E.~Witten,
  Adv.\ Theor.\ Math.\ Phys.\  {\bf 2}, 253 (1998)

\bibitem{Herzog:2006gh}
  C.~P.~Herzog, A.~Karch, P.~Kovtun, C.~Kozcaz and L.~G.~Yaffe,
  JHEP {\bf 0607}, 013 (2006)

\bibitem{Gubser:2006nz}
  S.~S.~Gubser,
  Nucl.\ Phys.\ B {\bf 790}, 175 (2008)

\bibitem{CasalderreySolana:2007qw}
  J.~Casalderrey-Solana and D.~Teaney,
  JHEP {\bf 0704} (2007) 039
\bibitem{HoyosBadajoz:2009pv}
  C.~Hoyos-Badajoz,
  JHEP {\bf 0909}, 068 (2009)
\bibitem{Gursoy:2010aa}
  U.~Gursoy, E.~Kiritsis, L.~Mazzanti and F.~Nitti,
  JHEP {\bf 1012}, 088 (2010)

\bibitem{Gubser:2006qh}
  S.~S.~Gubser,
  Phys.\ Rev.\ D {\bf 76}, 126003 (2007)

\bibitem{Gursoy:2007cb}
  U.~Gursoy and E.~Kiritsis,
  JHEP {\bf 0802}, 032 (2008)

\bibitem{Gursoy:2007er}
  U.~Gursoy, E.~Kiritsis and F.~Nitti,
  JHEP {\bf 0802}, 019 (2008)

\bibitem{Gubser:2008ny}
  S.~S.~Gubser and A.~Nellore,
  Phys.\ Rev.\ D {\bf 78}, 086007 (2008)

\bibitem{Gursoy:2009kk}
  U.~Gursoy, E.~Kiritsis, G.~Michalogiorgakis and F.~Nitti,
  JHEP {\bf 0912}, 056 (2009)

\bibitem{Abbasi:2012qz}
  N.~Abbasi and A.~Davody,
  JHEP {\bf 1206}, 065 (2012)

\bibitem{Kiritsis:2011bw}
  E.~Kiritsis, L.~Mazzanti and F.~Nitti,
  J.\ Phys.\ G G {\bf 39}, 054003 (2012)

\bibitem{pow} S. Frixione, P. Nason and G. Ridolfi, JHEP {\bf 09} 126 (2007)

\bibitem{expp} B. Abelev {\it et al.}  , ALICE Collaboration, JHEP {\bf 01} 128 (2012)   


\bibitem{setiii} C. Peterson {\it et al.},  Phys. Rev. D {\bf 27}, 105 (1983).

\bibitem{bjorken} J. D. Bjorken, Phys. Rev. D {\bf 27}, 140 (1983).



\bibitem{vic} A. K. Chaudhuri and V. Roy, Phys. Rev. C {\bf 84}, 027902 (2011)

\bibitem{eoss} U.~Gursoy, E.~Kiritsis, G.~Mazzanti and F.~Nitti, Nucl.\ Phys.\ B {\bf 820}, 148 (2009)


\bibitem{CasalderreySolana:2006rq}
  J.~Casalderrey-Solana and D.~Teaney,
  Phys.\ Rev.\ D {\bf 74}, 085012 (2006)

\bibitem{rad_con} K. B. Fadafan, H. Liu, K. Rajagopal and U. Wiedemann, Eur.Phys.J.C {\bf 61} 553 (2009)

\bibitem{rad_non} M. Ali-Akbari and U. Gursoy, arXiv:1110.5881 [hep-th]


\end{thebibliography}
\end{document}